\documentclass[%
reprint,
superscriptaddress,
showpacs,preprintnumbers,
 amsmath,amssymb,
 aps,
 longbibliography,
prb,
floatfix,
]{revtex4-1}
\usepackage[english]{babel}
\usepackage{multirow}
\usepackage[thinspace]{SIunits}
\usepackage{xspace}
\usepackage{graphicx}% Include figure files
\usepackage{dcolumn}% Align table columns on decimal point
\usepackage{bm}% bold math
\usepackage{color}
\usepackage{wrapfig}
\usepackage{hyperref}
\hypersetup{
    unicode=false,          % non-Latin characters in Acrobat’s bookmarks
    pdftoolbar=true,        % show Acrobat’s toolbar?
    pdfmenubar=true,        % show Acrobat’s menu?
    pdffitwindow=false,     % window fit to page when opened
    pdfstartview={FitH},    % fits the width of the page to the window
    pdftitle={My title},    % title
    pdfauthor={Author},     % author
    pdfsubject={Subject},   % subject of the document
    pdfcreator={Creator},   % creator of the document
    pdfproducer={Producer}, % producer of the document
    pdfkeywords={keyword1} {key2} {key3}, % list of keywords
    pdfnewwindow=true,      % links in new window
    colorlinks=true,       % false: boxed links; true: colored links
    linkcolor=blue,          % color of internal links (change box color with linkbordercolor)
    citecolor=blue,        % color of links to bibliography
    filecolor=blue,      % color of file links
    urlcolor=blue,       % color of external links
    plainpages=false        % hat Einfluss auf die Seitennummerierung
}

\newcommand{\etal}{\textit{et al.}\xspace}

\newcommand{\Tc}{$T_c$\xspace}

\newcommand{\BFCA}{\mbox{Ba(Fe$_{\mathrm{1-x}}$Co$_{\mathrm{x}}$)$_{2}$As$_2$}\xspace}

\newcommand{\BKFA}{\mbox{Ba$_{\mathrm{1-x}}$K$_\mathrm{x}$Fe$_{2}$As$_2$}\xspace}

\newcommand{\BFA}{\mbox{BaFe$_{2}$As$_2$}\xspace}
\newcommand{\FA}{\mbox{Fe$_{2}$As$_2$}\xspace}

\newcommand{\AFA}{\mbox{\textit{A}Fe$_{2}$As$_2$}\xspace}
\newcommand{\CKFA}{\mbox{CaKFe$_{4}$As$_4$}\xspace}

\newcommand{\Alg}{$\rm{A_{1g}}$\xspace}

\newcommand{\AZg}{$\rm{A_{2g}}$\xspace}
\newcommand{\Blg}{$\rm{B_{1g}}$\xspace}
\newcommand{\BZg}{$\rm{B_{2g}}$\xspace}

\newcommand{\Eg}{$\rm{E_{g}}$\xspace}
\newcommand{\grd}{$^{\circ}$\xspace}

\newcommand{\wavenumb}{\mathrm{cm}^{-1}}

\begin{document}

%\title{Lattice dynamics and phase transitions in detwinned \BFA}
\title{Subdominant $d$-wave interaction in superconducting \CKFA ?}

\date{\today}

\author{D. Jost}
\affiliation{Walther Meissner Institut, Bayerische Akademie der Wissenschaften,
85748 Garching, Germany}
\affiliation{Fakult\"at f\"ur Physik E23, Technische Universit\"at M\"unchen,
85748 Garching, Germany}

\author{J.-R. Scholz}
\affiliation{Walther Meissner Institut, Bayerische Akademie der Wissenschaften,
85748 Garching, Germany}
\affiliation{Fakult\"at f\"ur Physik E23, Technische Universit\"at M\"unchen,
85748 Garching, Germany}

\author{U. Zweck}
\affiliation{Walther Meissner Institut, Bayerische Akademie der Wissenschaften,
85748 Garching, Germany}
\affiliation{Fakult\"at f\"ur Physik E23, Technische Universit\"at M\"unchen,
85748 Garching, Germany}

\author{W. R. Meier}
\affiliation{Department of Physics and Astronomy, Iowa State University, Ames, Iowa 50011, USA}
\affiliation{Division of Materials Science and Engineering, Ames Laboratory, Ames, Iowa 50011, USA}

\author{A.\,E.~B\"ohmer}
\altaffiliation{Present address: Karlsruher Institut f\"ur Technologie, Institut f\"ur Festk\"orperphysik, 76021 Karlsruhe, Germany}
\affiliation{Division of Materials Science and Engineering, Ames Laboratory, Ames, Iowa 50011, USA}

\author{P. C. Canfield}
\affiliation{Division of Materials Science and Engineering, Ames Laboratory, Ames, Iowa 50011, USA}
\affiliation{Department of Physics and Astronomy, Iowa State University, Ames, Iowa 50011, USA}

\author{N. Lazarevi\'c}
\affiliation{Center for Solid State Physics and New Materials, Institute of Physics Belgrade, University of Belgrade, Pregrevica 118, 11080 Belgrade, Serbia}

\author{R. Hackl}
\affiliation{Walther Meissner Institut, Bayerische Akademie der Wissenschaften,
85748 Garching, Germany}

%\affiliation{Walther Meissner Institute, Bavarian Academy of Sciences and
%Humanities, 85748 Garching, Germany}
%\email{hackl@wmi.badw.de}

%------------------------------------------------------------------------------------------------------------
\begin{abstract}
  We report inelastic light scattering results on the stoichiometric and fully ordered superconductor \CKFA as a function of temperature and light polarization. In the energy range between 10 and 315\,cm$^{-1}$ (1.24 and 39.1\,meV) we observe the particle-hole continuum above and below the superconducting transition temperature $T_c$ and 7 of the 8 Raman active phonons. The main focus is placed on the analysis of the electronic excitations. Below $T_c$ all three symmetries projected with in-plane polarizations display a redistribution of spectral weight characteristic for superconductivity. The energies of the pair-breaking peaks in  \Alg and \BZg symmetry are in approximate agreement with the results from photoemission studies. In \Blg symmetry the difference between normal and superconducting state is most pronounced, and the feature is shifted downwards with respect to those in \Alg and \BZg symmetry. The maximum peaking at 134\,cm$^{-1}$ (16.6\,meV) has a substructure on the high-energy side. We interpret the peak at 134\,cm$^{-1}$ in terms of a collective Bardasis-Schrieffer (BS) mode and the substructure as a remainder of the pair-breaking feature on the electron bands. There is a very weak peak at 50\,cm$^{-1}$ (6.2\,meV) which is tentatively assigned to another BS mode.

\end{abstract}
%------------------------------------------------------------------------------------------------------------
\maketitle
%------------------------------------------------------------------------------------------------------------
%------------------------------------------------------------------------------------------------------------
\section{Introduction}

\begin{figure}[h!]
  \centering
  \includegraphics[width=7.5cm]{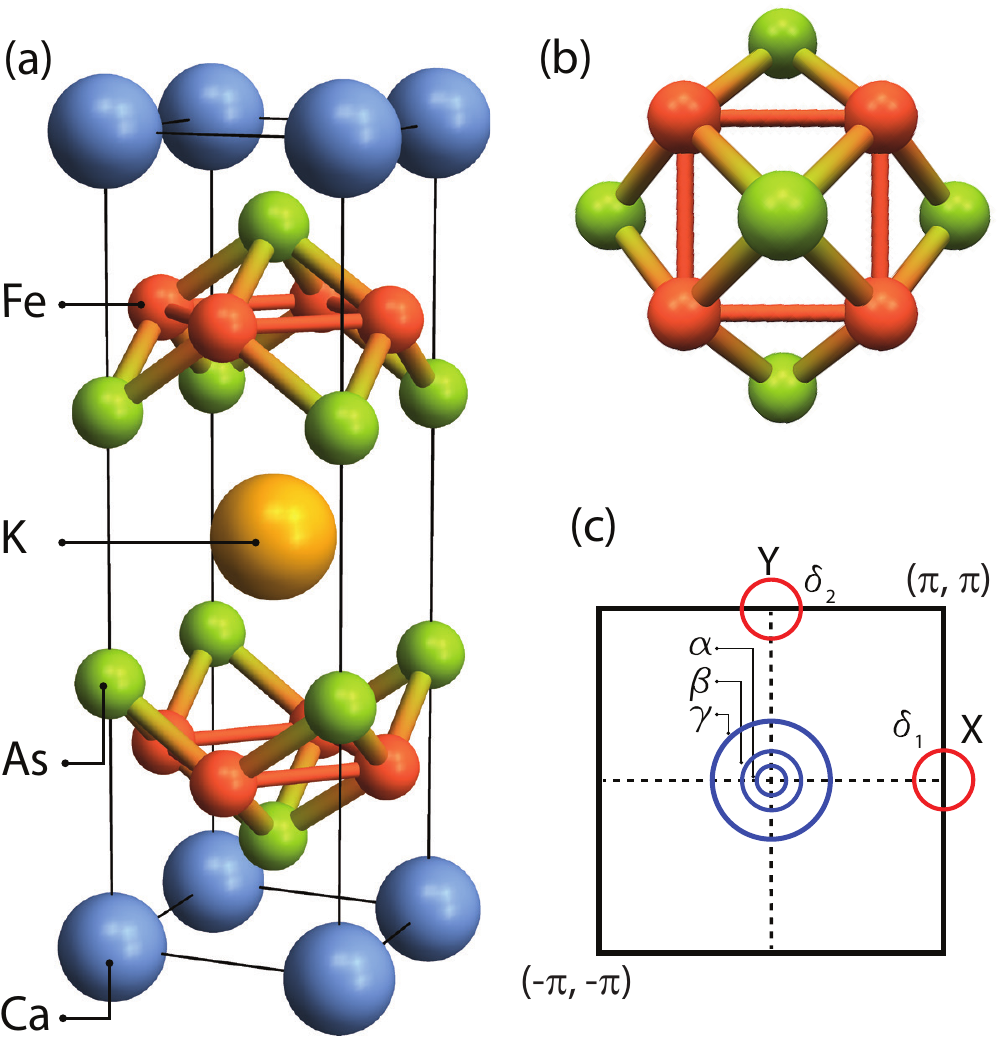}
  \caption{Structure and Fermi surface of \CKFA. (a) Tetragonal unit cell with the Ca, K, Fe, and As atoms shown in blue, gold,  red, and respectively, green \cite{Iyo:2016}. (b) Quasi 2D \FA layer with the full orange line depicting the 1\,Fe unit cell. (c) Brillouin zone of the 1\,Fe cell. The three hole bands labeled $\alpha$, $\beta$, and $\gamma$ (blue) encircle the $\Gamma$ point.  The electron bands $\delta_1$ and $\delta_2$ (red) are centered at the $X$ and $Y$ point, respectively [adopted from Mou \etal \cite{Mou:2016}].
  }
  \label{fig:CKFA}
\end{figure}

\CKFA is among the few iron-based compounds which are superconducting at high transition temperature \Tc at stoichiometry \cite{Iyo:2016} since the Ca and K atoms form alternating intact layers as shown in Fig.~\ref{fig:CKFA}. The high degree of order allows one to get as close to the intrinsic properties of the material class as possible since the effects of disorder are expected to be negligible or at least significantly smaller than in solid solutions such as \BKFA or \BFCA. For instance the residual resistivity ratio (RRR) reaches 15 and is much  higher than for \BFCA and comparable or better than for \BKFA \cite{Meier:2016}. Other transport and the thermodynamic properties \cite{Meier:2016} highlight the similarities to optimally or overdoped \BKFA.

These similarities include the electronic structure, in particular the Fermi surfaces [Fig.~\ref{fig:CKFA}\,(c)] and the superconducting energy gaps \cite{Mou:2016}. The gaps were found to be rather isotropic on the individual bands having values of $2\Delta_{\alpha} = 21\ \mathrm{meV}$, $2\Delta_{\beta}=24 \ \mathrm{meV}$, $2\Delta_{\gamma} = 16 \ \mathrm{meV}$ and $2\Delta_\delta = 24 \ \mathrm{meV}$ on the three hole bands ($\alpha$, $\beta$, $\gamma$) and the electron band ($\delta$), respectively. The good nesting observed between the $\beta$ and $\delta$ bands was considered to support $s$-wave interband pairing \cite{Mou:2016,Biswas:2017,Yang:2017} as proposed earlier for the iron-based materials in general \cite{Mazin:2008}.
%%%%%%%%%%%%%%%%%%%%%%%%%%%%%%%%%%%%%%%%%%%%%%%%%%%%%%%%%%%%%
%%%%%%%%%%%%%%%%%%%%%%%%%%%%%%%%%%%%%%%%%%%%%%%%%%%%%%%%%%%

Since there is also nesting among the electron bands one can expect a competing pairing interaction with $d$-wave symmetry as a subleading instability \cite{Hirschfeld:2016}. Indications thereof were found recently in Raman scattering experiments on \BKFA \cite{Kretzschmar:2013,Bohm:2014,Bohm:2017}. The subleading channel manifests itself as a narrow line below the gap edge given that the gap is clean as, e.g., in \BKFA \cite{Bardasis:1961,Monien:1990,Scalapino:2009}. The bound state has its origin in a phase fluctuation of the  condensate of Cooper pairs. The experimental identification of this excitation as a Bardasis-Schrieffer (BS) exciton rests on the shape of the line, its temperature dependence and the spectral weight transfer from the pair-breaking feature to the in-gap mode as described in detail in Refs. \onlinecite{Bohm:2014,Bohm:2017}.

Given its nearly clean gap and the high crystal quality, \CKFA is an excellent candidate for scrutinizing the superconducting properties of hole doped 122 systems. We aim at answering the question as to wether or not subleading channels can also be observed in systems other than \BKFA and can be identified as generic.

%\CKFA is interesting due to the sole occurrence of superconductivity, as this allows to study the bare Cooper pairing without interference of other competing phases such as magnetic ordering. Additionally, the highly ordered single crystals and their cleavable surfaces are very suitable for polarisation resolved Raman scattering studies.

\section{Experiment}
\label{sec:experiment}

Calibrated Raman scattering equipment was used for the experiments. The sample was attached to the cold finger of a He-flow cryostat. For excitation a diode-pumped solid state laser emitting at 575\,nm (Coherent GENESIS MX-SLM 577-500) was used.
The polarization of the incoming light was adjusted in a way that the light inside the sample had the proper polarization state. The absorbed power (inside the sample) was set at  typically $P_a=2$\,mW independent of polarization. By setting the polarizations of the incident and scattered photons the four symmetries \Alg, \AZg, \Blg, and \BZg of the D\textsubscript{4h} space group can be accessed. For the symmetry assignment we use the 1\,Fe unit cell [see Fig.~\ref{fig:CKFA}\,(b) and (c)] since the density of states at the Fermi energy $E_{\rm F}$ is nearly entirely derived from Fe orbitals. The related projections in the first Brillouin zone (BZ) are visualized in Fig.~\ref{Afig:SR} of Appendix \ref{sec:SR}. In this work the focus is placed on low energies, where \AZg is negligibly small. The polarization combinations $RR$, $xy$, and $x^\prime y^\prime$ almost exclusively project the \Alg, \BZg, and \Blg symmetries as desired. Here $x$ and $y$ are horizontal and vertical, respectively, in the laboratory system. Note that the out-of-phase vibration of the Fe atoms appears in \BZg symmetry in the 1\,Fe unit cell rather than in \Blg symmetry of the crystallographic unit cell hosting two Fe atoms per Fe$_2$As$_2$ plane. For more details see section \ref{Asec:phonons}. The spectra are represented as response functions $R\chi^{\prime\prime}(T), \Omega )$ which are obtained by dividing the measured cross section by the thermal Bose factor $\{1+n(\Omega, T)\} = \{1+[\exp(\hbar\omega/k_BT)-1]^{-1}\}$. $R$ is an experimental constant.

The \CKFA single crystals were grown from the FeAs flux and characterized thoroughly as described by Meier and coworkers \cite{Meier:2016,Meier:2017}. The \Tc value of $35.21\pm0.10$\,K we found here is in the range $35.0\pm0.2$\,K determined by Meier \etal (see Fig. \ref{sec:Tc} in Appendix \ref{Afig:Tc}). The crystal structure of \CKFA is very similar to that of \AFA systems [cf. Fig. \ref{fig:CKFA}\,(a)] and belongs to the tetragonal $D_{4h}$ space group. Since \CKFA has alternating Ca and K planes between the Fe$_2$As$_2$ layers the point group is simple-tetragonal ($P4/mmm$) \cite{Iyo:2016} rather than the  body-centered tetragonal ($I4/mmm$) as \BFA [\onlinecite{Rotter:2008}].

\section{Results}
\label{sec:results}

\begin{figure}[ht]
  \centering
  \includegraphics[width=8.5cm]{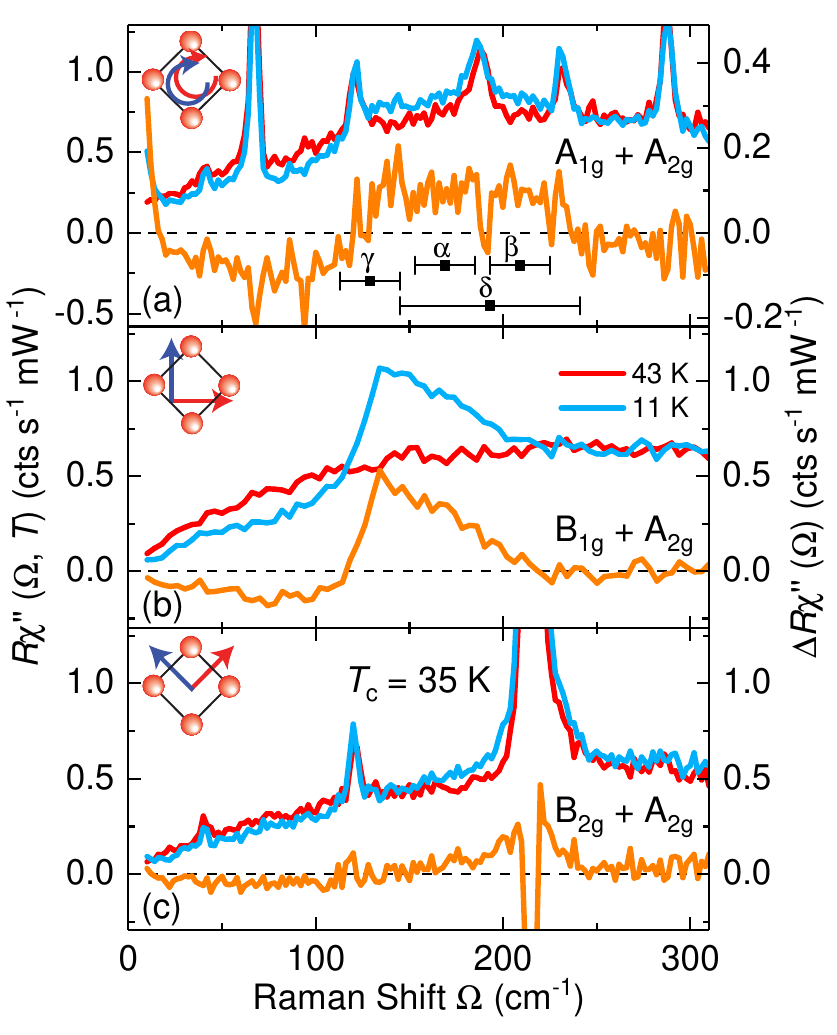}
  \caption{Raman response in \CKFA at symmetries and polarizations as indicated. Shown are raw data for $T\ll T_c$ (blue), $T>T_c$ (red) and difference spectra (orange). Phonon modes are present in the \Alg and \BZg spectra. (a)  The pair-breaking maximum extends from $\Omega_\mathrm{0}^\mathrm{(A{1g})} = 120\pm2\,\wavenumb$ to $\Omega_\mathrm{m} \approx 230\,\wavenumb$. The gap energies $2\Delta_i$ for the four bands $i$ observed by ARPES \cite{Mou:2016} are reproduced as horizontal bars. (b) The \Blg pair-breaking peak is well defined and sets on at $\Omega_\mathrm{0}^\mathrm{(B{1g})}=116\pm5\,\wavenumb$. The dashed line at low energies indicates the expected spectral dependence for a clean gap. (c) The weak \BZg pair-breaking peak is located slightly above $2\Delta_\beta$. The intersection of the normal and superconducting spectra is close to  $\Omega_\mathrm{0}^\mathrm{(B{2g})}=130\pm5\,\wavenumb$.
  }
  \label{fig:data-T=0}
\end{figure}

Fig.~\ref{fig:data-T=0} shows the normal (red) and superconducting (blue) Raman spectra of \CKFA at the three  polarization configurations (a) $RR$, (b) $xy$, and (c) $x^\prime y^\prime$. The sample is rotated by 45\grd with respect to the orientation in Figs.~\ref{fig:CKFA} and \ref{Afig:SR} in order to suppress any $c$-axis projection in the \Blg spectra ($xy$ in the laboratory system). Superimposed on the particle-hole continua we observe six and four phonons in $RR$ and $x^\prime y^\prime$ polarization, respectively \cite{Scholz:2017} which will be discussed in Appendix \ref{Asec:phonons}. As intended there are no phonons in $xy$ configuration, and the spectrum in the normal state is completely smooth to within the experimental error. The structure-less shape indicates that there is no polarization leakage and, more importantly, that there is no defect-induced scattering from phonons highlighting the high crystal quality.
%%%%%%%%%%%%%%%%%%%%%%%%%%%%%%%%%%%%%%%%%%%%%%%%%%%%%%%%%%%%%
%\onecolumngrid
%\begin{widetext}
%\end{widetext}
%\twocolumngrid
%%%%%%%%%%%%%%%%%%%%%%%%%%%%%%%%%%%%%%%%%%%%%%%%%%%%%%%%%%
We focus now exclusively on the electronic continua.

To this end we also plot the difference spectra,
\begin{equation}
  \Delta R \chi^{\prime\prime} (\Omega) = R\chi^{\prime\prime}(T = 11 \,\mathrm{K}, \Omega ) - R \chi^{\prime\prime}(T= 43\,\mathrm{K}),
  \label{eqn:differencespectrum}
\end{equation}
(orange in Fig.~\ref{fig:data-T=0}) along with the raw data of each polarization configuration. In the difference spectra $\Delta R \chi^{\prime\prime} (\Omega)$ temperature-independent structures such as (most of) the phonons and the presumably weak and temperature-independent luminescence contributions are eliminated. In this way the changes induced by superconductivity are highlighted.

All spectra show the typical changes upon entering the superconducting state: (i) The opening of the gap induces a suppression of the intensity below a cross-over energy of $\Omega_0=125, ~115, ~{\rm and}~ 130\,\mathrm{cm}^{-1}$ for \Alg, \Blg, and \BZg symmetry, respectively. In this range $\Delta R \chi^{\prime\prime} (\Omega)$ (orange) is negative. (ii) The intensity piles up above $\Omega_0$ due to a coherent superposition of pair-breaking and Bogoliubov quasiparticle excitations across the gap $2\Delta$. The amplitude of the redistribution is small in \Alg and \BZg symmetry [Fig.~\ref{fig:data-T=0}\,(a) and (c) and Figs.~\ref{Afig:A1g} and \ref{Afig:B2g}] but pronounced in \Blg symmetry. [Fig.~\ref{fig:data-T=0}\,(b) and Fig.~\ref{Afig:B1g}]. In the \Alg response [Fig.~\ref{fig:data-T=0}\,(a)] the signal at $\Omega \to 0$ is enhanced because of surface layers of accumulating residual gas molecules at low temperature (see Fig.~\ref{Afig:A1g}) and the insufficient suppression of the elastically scattered light in the case of parallel light polarizations ($RR$ here).

The first striking observation is the nearly symmetry-independent cross-over energy $\Omega_0$ where the normal (red) and superconducting (blue) spectra intersect each other or where $\Delta R \chi^{\prime\prime} (\Omega)$ changes sign. Yet, the intensity for $\Omega<\Omega_0$ does not vanish entirely as expected for a clean gap but is only reduced. No additional structures are observed in the \Alg and \BZg spectra while a weak hump appears at approximately 50\,cm$^{-1}$ [see Fig.~\ref{fig:data-T=0}\,(b) and arrows in Fig.~\ref{fig:Dchi-B1g-T}\,(b) to (f)] in \Blg symmetry. It appears also at elevated temperatures as shown in Fig.~\ref{Afig:B1g} and is therefore considered a robust feature.

Second,  whereas the normal and the superconducting spectra merge at similar energies close to $\Omega_{\rm m}=230$\,cm$^{-1}$ in all symmetries the distribution of spectral weight in the range $\Omega_0<\Omega<\Omega_{\rm m}$ varies substantially. In none of the symmetries the pair-breaking features display the typical shape. The pair-breaking maximum in \BZg symmetry is found at approximately 215\,cm$^{-1}$ right underneath the Fe phonon. The negative intensity at 215\,cm$^{-1}$ shows that the phonon is renormalized below $T_c$, and an influence of this renormalization on the electronic features cannot be excluded. However, the gap below 130\,cm$^{-1}$ indicates the presence of an intensity redistribution below $T_c$. In \Alg symmetry a wide plateau is observed between $\Omega_0$ and $\Omega_{\rm m}$. Finally in \Blg symmetry, a pronounced peak is found at 135\,cm$^{-1}$ above which the intensity decays. Upon studying various temperatures a secondary maximum at about 165\,cm$^{-1}$ can be resolved as shown in Fig.~\ref{Afig:B1g} in the .

For the discussion below we additionally plot in Fig.~\ref{fig:data-T=0}\,(a) the gap energies $2\Delta_i$ as horizontal bars according to a recent photoemission study \cite{Mou:2016}, where $i$ is the band index [c.f. Fig.~\ref{fig:CKFA}\,(c)]. The width of the bars corresponds to the error bars of order $\pm10$\% indicated there.

%%%%%%%%%%%%%%%%%%%%%%%%%%%%%%%%%%%%%%%%%%%%%%%%%%%%%%%%%%%%%
\section{Discussion}
The main purpose of this section is a balanced discussion of the possible interpretations of the electronic Raman spectra presented in section \ref{sec:results}. Can the spectra be interpreted exclusively in terms of pair-breaking or are collective modes, similarly as in \BKFA, necessary for a more consistent explanation?
\subsection{Gap energies}
\label{sec:energies}
Using yellow excitation we find a strong redistribution of spectral weight in \Blg symmetry similarly as in a simultaneous Raman study using red photons \cite{Zhang:2018} (Note that \Blg and \BZg are interchanged in these studies.) With yellow photons the redistribution can be observed in all three symmetry channels. As already noticed earlier this difference in the experimental results may be traced back to orbital dependent resonance effects \cite{Bohm:2017}.

The highest pair-breaking energy in our study is observed in \BZg symmetry implying a maximal gap energy of $2\Delta_{\rm max} \approx 215$\,cm$^{-1}$. %(Note that Zhang \etal use the crystallographic unit cell for the symmetry assignment. Thus \Blg and \BZg are interchanged.)
This energy corresponds to $\Delta_{\rm max}=13.3$\,meV slightly higher than the largest gaps derived for the $\beta$ and $\delta$ bands, $\Delta_{\beta,\delta}=12$\,meV using angle-resolved photoemission spectroscopy (ARPES) \cite{Mou:2016}. 215\,cm$^{-1}$ coincides with the edge of the \Alg pair-breaking feature [see Fig.~\ref{fig:data-T=0}], and we conclude that the ARPES data slightly underestimate the gap energies found by Raman scattering as already observed for \BKFA [\onlinecite{Kretzschmar:2013,Bohm:2014}]. Similarly, the lowest gap energy of $\Delta_{\gamma}=8$\,meV is below 9.3\,meV expected from the lower edge in the \Alg spectra. There are no structures in the \Alg and \BZg spectra which one could associate with the gap energy on the $\alpha$ band, $\Delta_{\alpha}=10.5$\,meV, obtained from ARPES. On this basis we conclude that the maximal gap energies derived from the \Alg and \BZg Raman spectra are in the same range of approximately $9\,k_BT_c$ as in \BKFA close to optimal doping \cite{Bohm:2014}.

The question arises whether ARPES and Raman results are compatible with the selection rules. As shown in Fig.~\ref{Afig:SR} all bands should be visible in \Alg symmetry with comparable weight upon neglecting resonance effects. In fact, all energies are represented in the spectra shown in Fig.~\ref{fig:data-T=0}\,(a). Even if a Leggett mode contributes to the \Alg spectra, as suggested recently \cite{Cea:2016}, this conclusion survives since the Leggett modes are expected close to the maximal gap energies in the Fe-based systems. The \BZg spectra are less easily to be reconciled  with this scenario since the gaps on the hole bands should be projected with a similar spectral weight as that of the electron band. Yet we find only a contribution from the largest gap. Although the overall intensity is very weak here the absence of contributions from the $\gamma$ band cannot be explained by the variation of the peak height with $|\Delta|^2$ [\onlinecite{Devereaux:2007}] or by applying the symmetry selection rules. Either a phenomenological treatment as for \BKFA [\onlinecite{Bohm:2014}] or a detailed resonance study needs to be performed which, however, is beyond the scope of this work.

Given that the single particle gap energies are by and large reproduced in the \Alg and \BZg spectra it is important to understand the \Blg spectra. As shown in Fig.~\ref{fig:data-T=0}\,(c) the energies appearing there are well below those of the \Alg and \BZg spectra. This is particularly surprising as the $\delta$ bands are expected to be projected fully (and not marginally) in \Blg symmetry (see Fig.~\ref{Afig:SR}) as opposed to all hole bands. Thus, the argument that the strongest peak in the \Blg spectra results from the $\gamma$ band can be discarded.

\subsection{\Blg Response}
\label{sec:B1g-finalstate}

\begin{figure}
  \includegraphics[width=7cm]{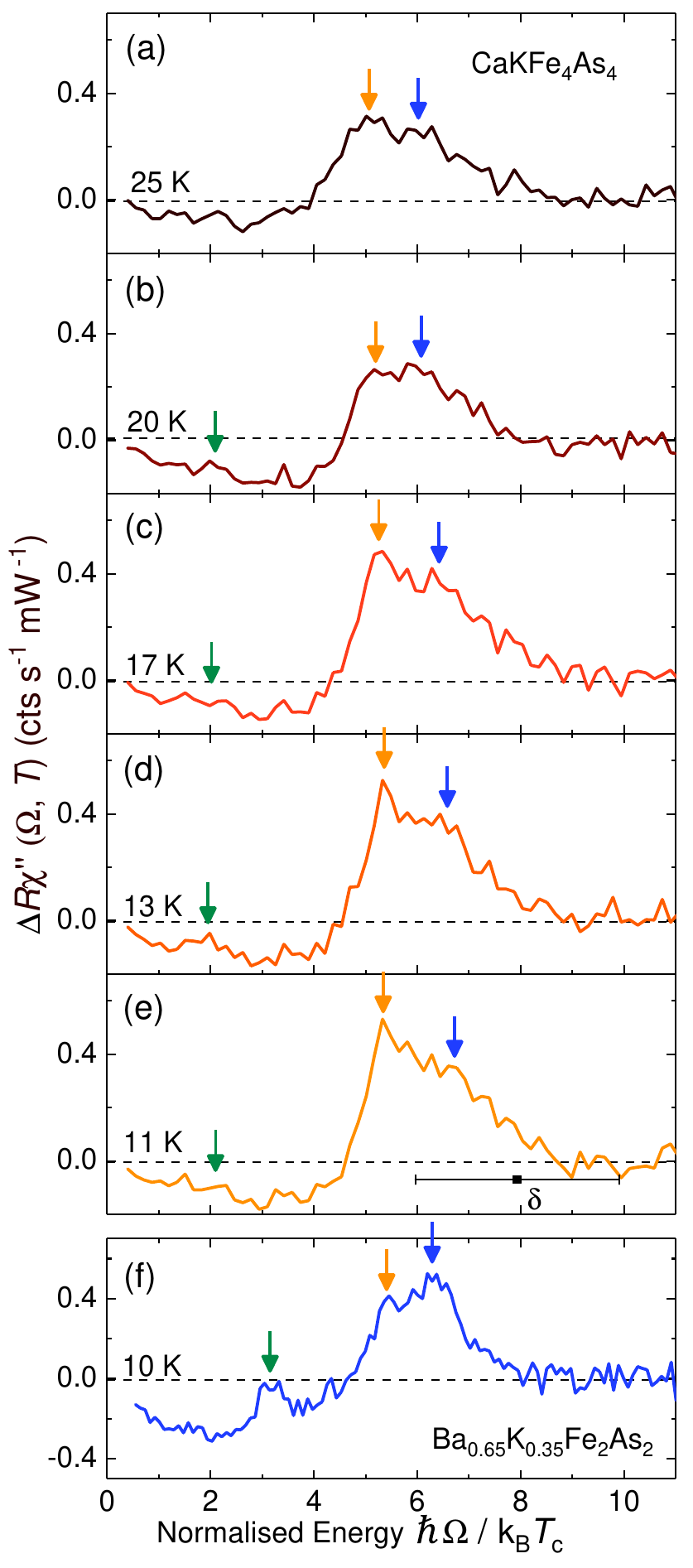}
  \caption{Difference spectra of the \Blg Raman response for temperatures as indicated. The main peak exhibits a double structure (orange and blue arrows). A second hump is visible from 25\,K down to 13\,K (green arrow, see also the raw data in Fig.~\ref{fig:data-T=0}). (f) Difference spectrum of Ba$_{0.65}$K$_{0.35}$As$_2$Fe$_2$. From \cite{Bohm:2017}. {The arrows show two Bardasis-Schrieffer modes at 3.1 (green arrow) and 5.2\,$k_BT_c$ (orange arrow). The remainder of the pair breaking peak is located at 6.2\,$k_BT_c$ (blue arrow) since the high-energy part is drained into the BS modes.}
  }
  \label{fig:Dchi-B1g-T}
\end{figure}

There is only one alternative scenario which reconciles the results observed in the three Raman active symmetries and the ARPES results: The \Blg spectra do not directly reflect gap energies but rather are shifted downward by final state interaction as discussed for \BKFA in earlier work \cite{Kretzschmar:2013,Bohm:2014,Bohm:2017}. The similarity can be observed directly by comparing the data in Fig.~\ref{fig:Dchi-B1g-T}. The difference spectra as a function of temperature indicate that the \Blg peak has a robust shoulder on the high-energy side. The overall shape is surprisingly similar to the spectra of $\mathrm{Ba_{0.65}K_{0.35}Fe_2As_2}$.

Following this reasoning we identify the maximum of the \Blg spectra at 135\,cm$^{-1}$ with a collective mode pulled off the maximal gap energy on the $\delta$ band due to a $d_{x^2-y^2}$ wave subleading interaction among the two electron bands predicted theoretically \cite{Scalapino:2009} and observed in \BKFA. The hump at approximately 165\,cm$^{-1}$ is then the remaining intensity of the pair-breaking peak on the $\delta$ band after switching on the final state interaction which induces a transfer of intensity from the pair-breaking peak into the bound state \cite{Monien:1990,Scalapino:2009,Bohm:2014}. Only in this way the missing intensity in the range of $2\Delta_\delta$ can be explained consistently.

It is tempting to explain the faint peak close to 50\,cm$^{-1}$ ($2\,k_BT_c$) in terms of a second BS mode in a similar fashion as in \BKFA \cite{Bohm:2017}. This would mean that the subdominant coupling is already very strong, and \CKFA is on the brink of a $d$-wave instability. The very weak intensity of the peak argues in this direction since the BS mode is expected to vanish when $d$-wave pairing wins. Yet, the vanishingly small intensity is also the Achilles heel of the argumentation, and a robust statement is possible only on the basis of a microscopic model, which includes the derivation of the eigenvectors of the subdominant pairing channels, as proposed for \BKFA \cite{Bohm:2017}. Such an expensive theoretical treatment is beyond the scope of this experimental study.

%BKFA\cite{Xu2008_EPL_BKFA, Zabolotnyy2009}.

\section{Conclusion}
We investigated the recently discovered superconductor \CKFA with inelastic light scattering as a function of photon polarisation and temperature. Using yellow light (575\,nm) superconducting features were found in \Alg, \Blg and \BZg symmetry.

A weak but well-defined pair-breaking feature is found at 215\,cm$^{-1}$ (corresponding to $\Delta=13.3$\,meV) in \BZg symmetry which is slightly above the largest gaps observed by ARPES for the $\beta$ and the $\delta$ bands \cite{Mou:2016} and close to the energy $\Omega_{\rm m}$ where the normal and the superconducting spectra merge in all symmetry projections. This feature is also present in the \Alg spectra. In addition to the high-energy feature the \Alg intensity displays a plateau-like shape down to $\Omega_0^{\rm (A1g)}=125$\,cm$^{-1}$. Given the small discrepancies between the gap energies derived from the ARPES data and the Raman spectra one can conclude that the \Alg spectra reflect the entire range of gap energies of \CKFA even though the individual gap energies cannot be resolved.

In \Blg symmetry, the superconducting feature is centered at lower energy than in the two other symmetries. We interpret the sharp maximum at 135\,cm$^{-1}$ as a collective Bardasis-Schrieffer mode pulled off the maximal gap on the $\delta$ band similarly as in the sister compound $\mathrm{Ba_{0.65}K_{0.35}Fe_2As_2}$. The shoulder at approximately 165\,cm$^{-1}$ is a remainder of the pair-breaking peak losing most of its intensity to the collective mode \cite{Scalapino:2009,Bohm:2014}. Whether or not the weak structure at 50\,cm$^{-1}$ is another BS mode with even stronger coupling cannot be decided with certainty because of the fading intensity. If this interpretation could be supported further \CKFA would be closer to a $d$-wave instability than \BKFA. The smaller \Tc of \CKFA argues in this direction since a strong $d$ pairing channel frustrates the $s$-wave ground state and reduces \Tc. Even without dwelling on the peak at 50\,cm$^{-1}$ we may conclude that \CKFA is a true sibling of \BKFA \cite{Bohm:2017} thus demonstrating that pairing fingerprints can be observed preferably in materials with clean gaps.

\section*{Acknowledgements}
We acknowledge valuable discussions with G. Blumberg. The work was supported by the Friedrich-Ebert-Stiftung, the Transregional Collaborative Research Center TRR\,80, and the Serbian Ministry of Education, Science and Technological Development under Project III45018.
We acknowledge support by the DAAD through the bilateral project between Serbia and Germany (grant numbers 56267076 and 57142964).
Work at Ames Laboratory was supported by the U.S. Department of Energy, Office of Basic Energy Sciences, Division of Materials Sciences and Engineering under Contract No. DE-AC02- 07CH11358.
W.R.M. was supported by the Gordon and Betty Moore Foundation’s EPiQS Initiative through Grant No. GBMF4411.

%\newpage
%\bibliographystyle{prsty}
%\bibliography{literature_ph}
%\bibliography{Lit_CKFA_DJ}
%\bibliography{D:/!papers/!bib/literatureR2}
%\bibliography{literatureR2}

%\clearpage
\renewcommand{\thefigure}{A\arabic{figure}}
\renewcommand{\theequation}{A\arabic{equation}}
\setcounter{figure}{0}
\setcounter{equation}{0}
\begin{appendix}
\label{sec:appendix}

\section{Raman Selection Rules}
\label{sec:SR}
\begin{figure}[ht]
  \centering
  \includegraphics[width=7.5cm]{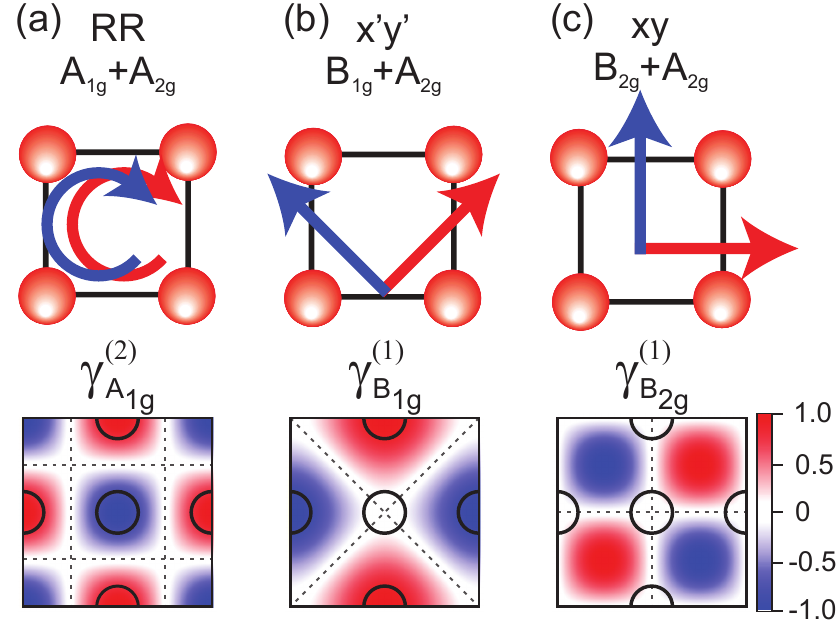}
  \caption{Raman form factors. The first row shows the polarizations w.r.t. the Fe sublattice. The second row displays the momentum dependences of the Raman vertices \cite{Devereaux:2007}. The \Alg vertex displayed here corresponds to $\gamma_{\rm A1g}^{(2)}$ which was found to be the most appropriate vertex for the FeAs systems \cite{Mazin:2010a,Bohm:2014} because of the opposite curvature of the electron and hole bands.
  }
  \label{Afig:SR}
\end{figure}

Using different combinations of light polarizations different parts of the Brillouin zone can be projected independently for particle-hole excitations \cite{Devereaux:1994}.
The weighting factors for the three Raman-active in-plane symmetries are shown in Fig.~\ref{Afig:SR}. Here the linear vertices are displayed which are derived entirely on the basis of symmetry arguments. In \Alg symmetry we use the second order vertex $\gamma_{\rm A1g}^{(2)}\propto \cos (k_x)\cos (k_y)$ which reflects the band structure better than the first order vertex because of the curvature of the hole and the electron bands \cite{Mazin:2010a,Bohm:2014}. (The zero-order vertex is a constant and entirely screened.)

For the experiments the samples were mounted on the cold finger in a way that the crystallographic axes match the laboratory system ($xy$). Because of the large angle of incidence, the incoming photons having polarization $x^\prime$ or $R$ have the same finite projection on the $c$-axis. Consequently, \Eg phonons, obeying $xz$ and $yz$ selection rules, are expected to appear in both $RR$ and $x^\prime y^\prime$ scattering configurations, whereas they are absent in $xy$ configuration.

%For the experiments the samples were mounted on the cold finger in a way that the crystallographic axes are horizontal ($\parallel ~x$) and vertical ($\parallel ~y$) in the laboratory system thus the Fe plane is rotated by 45\grd with respect to the first row  of Fig.~\ref{Afig:SR} (and Fig.~\ref{fig:CKFA}b and c). Then for $x^\prime y^\prime$ polarization (\BZg+\Alg symmetry) the polarization of the incoming photon ($x^\prime$) has the same net projection on the $c$-axis as for $R$. On the other hand, there is no $c$-axis projection in $xy$ polarization. Consequently, \Eg phonons appear only in $RR$ and $x^\prime y^\prime$ configurations for obeying $xz$ and $yz$ selection rules.
\section{Phonons}
\label{Asec:phonons}
\begin{figure}[b]
  \vspace{-3mm}
  \centering
  \includegraphics[width=7.5cm]{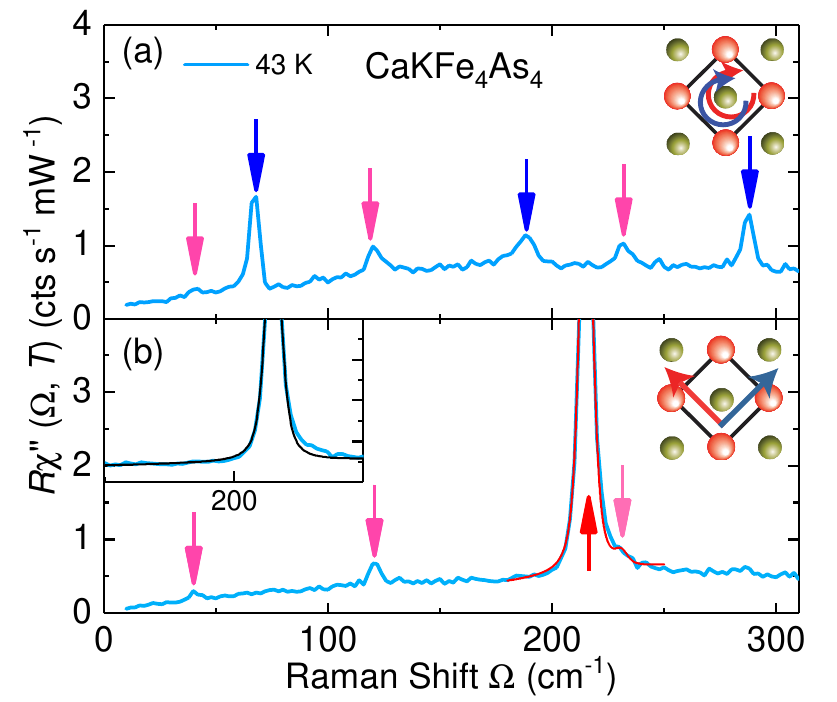}
  \caption{Raman-active phonons in \CKFA at 43\,K. The six and four lines in (a) $RR$ and, respectively, (b) $x^\prime y^\prime$ polarization are marked by blue, red, and magenta arrows for \Alg, \Blg, and \Eg symmetry, respectively. The orientation of the crystal with respect to the laboratory frame along with the polarization symbols is shown pictorially with Fe and As displayed in red and olive, respectively. Due to the large angle of incidence of 66\grd both polarization configurations have a finite projection on the $c$-axis of the crystal. Thus the \Eg phonons are seen with approximately the same intensity in $RR$ and $x^\prime y^\prime$. The \Eg mode at 232\,cm$^{-1}$ is hidden underneath the \Blg mode. The asymmetry of the \Blg line can be explained entirely by a superposition of two lines as shown in (b). The inset in (b) shows the data and a symmetric Lorentz line.
  }
  \vspace{-3mm}
  \label{Afig:phonons}
\end{figure}

For the phonons it is more appropriate to use the crystallographic 2\,Fe unit cell of \CKFA having $P4/mmm$ symmetry as shown in Fig.~\ref{fig:CKFA}. The orientation of the crystal in the experiment is displayed as an inset in Fig.~\ref{Afig:phonons}. The energies of all phonon lines observed are collected in Table~\ref{Atab:phonons}.

The strongest line that appears at  215\,cm$^{-1}$ in $x^\prime y^\prime$ polarization is the out-of-phase $c$-axis Fe vibration having the proper \Blg symmetry in the crystallographic (2\,Fe) unit cell.

\begin{table}%[tb]
  \caption{Phonon energies of \CKFA at 43\,K.
  }
  \label{Atab:phonons}
  \renewcommand{\arraystretch}{1.8}
  \begin{ruledtabular}
  \begin{tabular}{clccccccc}
  pol                 & unit      & $E_g^{(1)}$ & $A_{1g}^{(As1)}$ & $E_g^{(2)}$ & $A_{1g}^{(As2)}$ & $B_{1g}^{(Fe)}$ & $E_g^{(3)}$ & $A_{1g}^{(Fe)}$\\
  \hline
  $RR$                & cm$^{-1}$ & 42   & 66   & 120  & 188  &      & 232  & 288  \\
                      & meV       & 5.21 & 8.18 & 14.9 & 23.3 & 26.7 & 28.8 & 35.7   \\
  $x^\prime y^\prime$ & cm$^{-1}$ & 42   &      & 120  &      & 215  & 232  &      \\
  \end{tabular}
  \end{ruledtabular}
\end{table}

The in-phase $c$-axis vibrations of the As atoms appear at 66 and 188\,cm$^{-1}$ in the $RR$ spectra. The line at 288\,cm$^{-1}$ corresponds to the in-phase $c$-axis vibrations of the Fe atoms. Three modes appear simultaneously with approximately the same intensity in $RR$ and $x^\prime y^\prime$ polarization both of which having the same finite projection of about 30\% on the $c$-axis of the incoming photons inside the crystal and are tentatively identified as three of the four \Eg modes expected to be Raman-active in the $P4/mmm$ structure of \CKFA. They are projected in $zx$ and $zy$ configuration thus have only approximately 10\% of the full intensity. The mode at 232\,cm$^{-1}$ cannot be observed independently in the $x^\prime y^\prime$ spectra since it is too close to the \Blg mode. We show an approximate decomposition in Fig.~\ref{Afig:phonons}\,(b). All \Eg phonons are shear modes with the Fe and As atoms moving parallel to the FeAs planes with different eigenvectors. In total one expects four \Eg modes \cite{Scholz:2017}. The missing fourth mode is either too weak to be observable or outside the range we studied.

\section{Sample characterization}
\label{sec:Tc}

\begin{figure}[t]
  \centering
  \includegraphics[width=7.5cm]{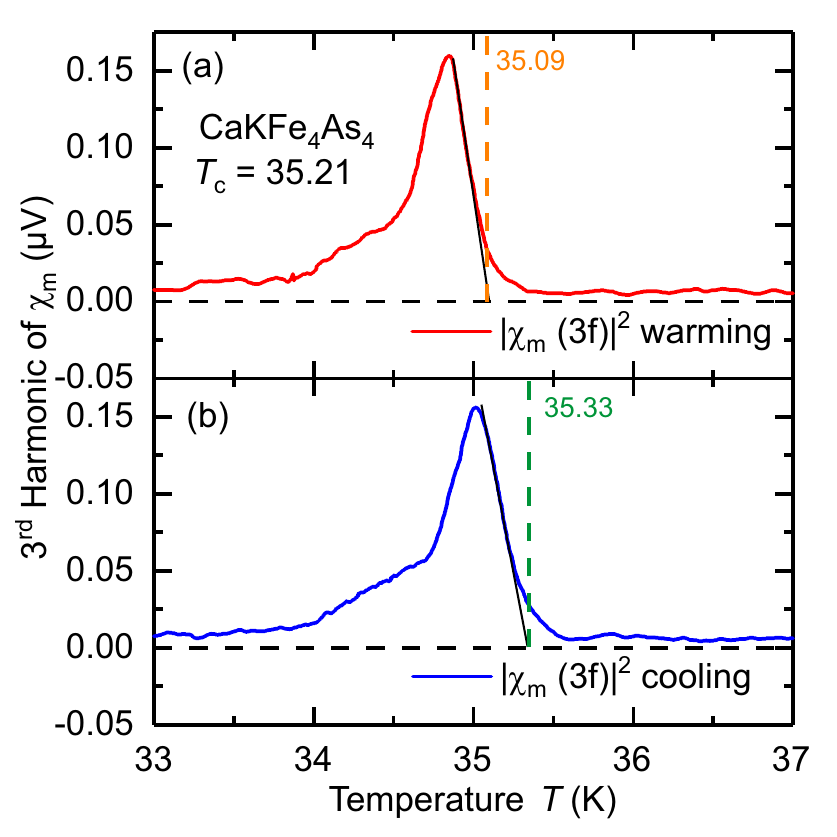}
  \caption{Nonlinear susceptibility of \CKFA. The two panels show the signal of the (a) heating and (b) cooling run. Since the sample lags slightly behind the sensor the indicated temperature is slightly higher and lower, respectively, between the curves. We derive $T_c=33.2$\,K and $\Delta T_c =0.3$\,K.
  }
  \label{Afig:Tc}
\end{figure}

The transition temperature \Tc of \CKFA sample used for the experiments was determined via the temperature dependence of the nonlinear magnetic susceptibility $\chi_{\rm m}^{(3)}(T)$ which is very sensitive for sample inhomogeneities \onlinecite{Raphael:1998}. Fig.  \ref{Afig:Tc} displays the results of the measurements on the samples used here for a cooling and a heating run.

As shown by Shatz and coworkers \cite{Shatz:1993} the $\chi_{\rm m}^{(3)}(T)$ curves displays a nearly linear onset right below \Tc given that there is no dc field, the amplitude of the exciting ac field is small (typically $\mu_0H_0<10^{-3}B_{\rm c1}$), and the superconducting transition is sharp ($\Delta T_c \to 0$). Below \Tc $\chi_{\rm m}^{(3)}(T)$ peaks at an amplitude-dependent temperature and decays again. The extrapolation to zero of the linear part between the peak and \Tc can be identified with the midpoint of the transition obtained from the linear susceptibility. If $\Delta T_c$ is not negligible $\chi_{\rm m}^{(3)}(T)$ has a foot above the extrapolated \Tc value. The width of the foot indicates $\Delta T_c$.

Upon measuring $\chi_{\rm m}^{(3)}(T)$ during cooling and heating with typical rates of $\pm1$\,K per min a small hysteresis of $0.1\dots0.2$\,K results. We use the average of the two individual \Tc values, $T_c=33.21$\,K, as the sample's transition temperature. $\Delta T_c =0.3$\,K can be derived from the foot of $\chi_{\rm m}^{(3)}(T)$ above \Tc.  The transition temperature is identical to that of Ref.\cite{Meier:2016}. We cannot identify any secondary phases.

\section{Raw Data}
\label{Asec:raw}
\begin{figure}[ht]
  \centering
  \includegraphics[width=7.5cm]{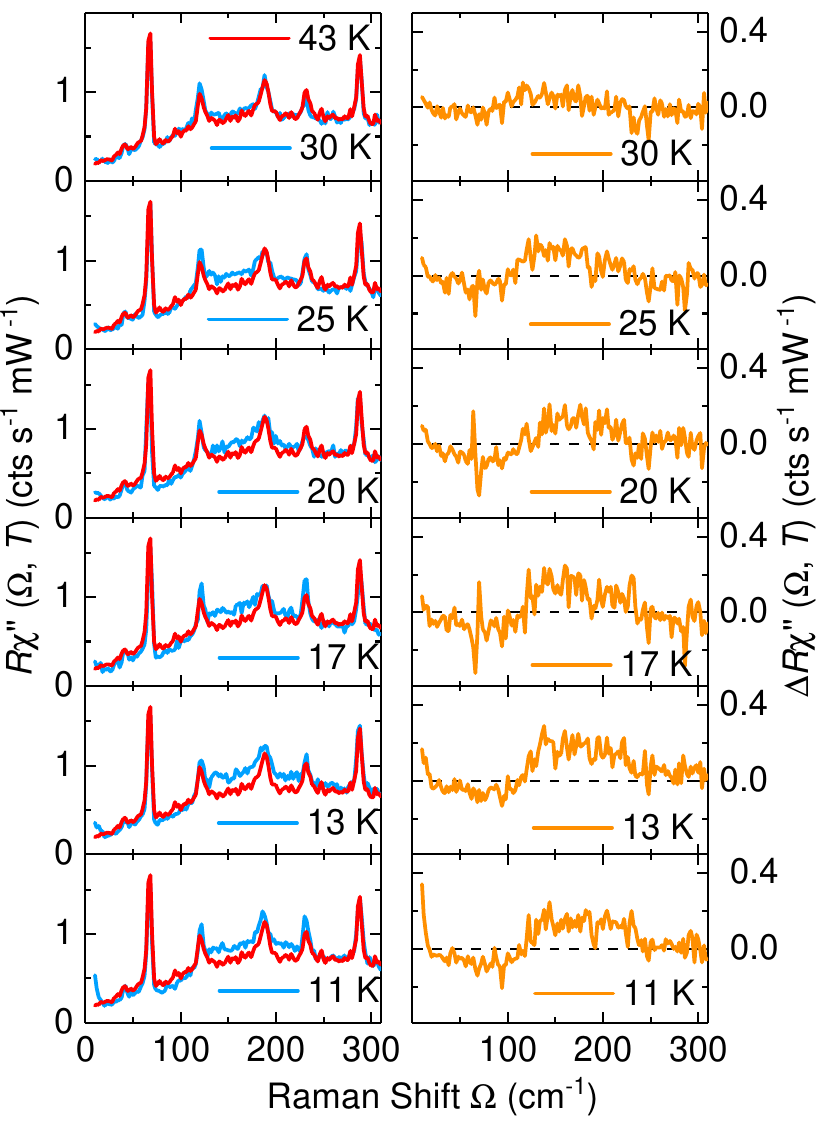}
  \caption{\Alg response of \CKFA for 43\,K and temperatures below \Tc as indicated. The raw response is shown on the left, the difference spectra are on the right.
  }
  \label{Afig:A1g}
\end{figure}

\begin{figure}[ht]
  \centering
  \includegraphics[width=7.5cm]{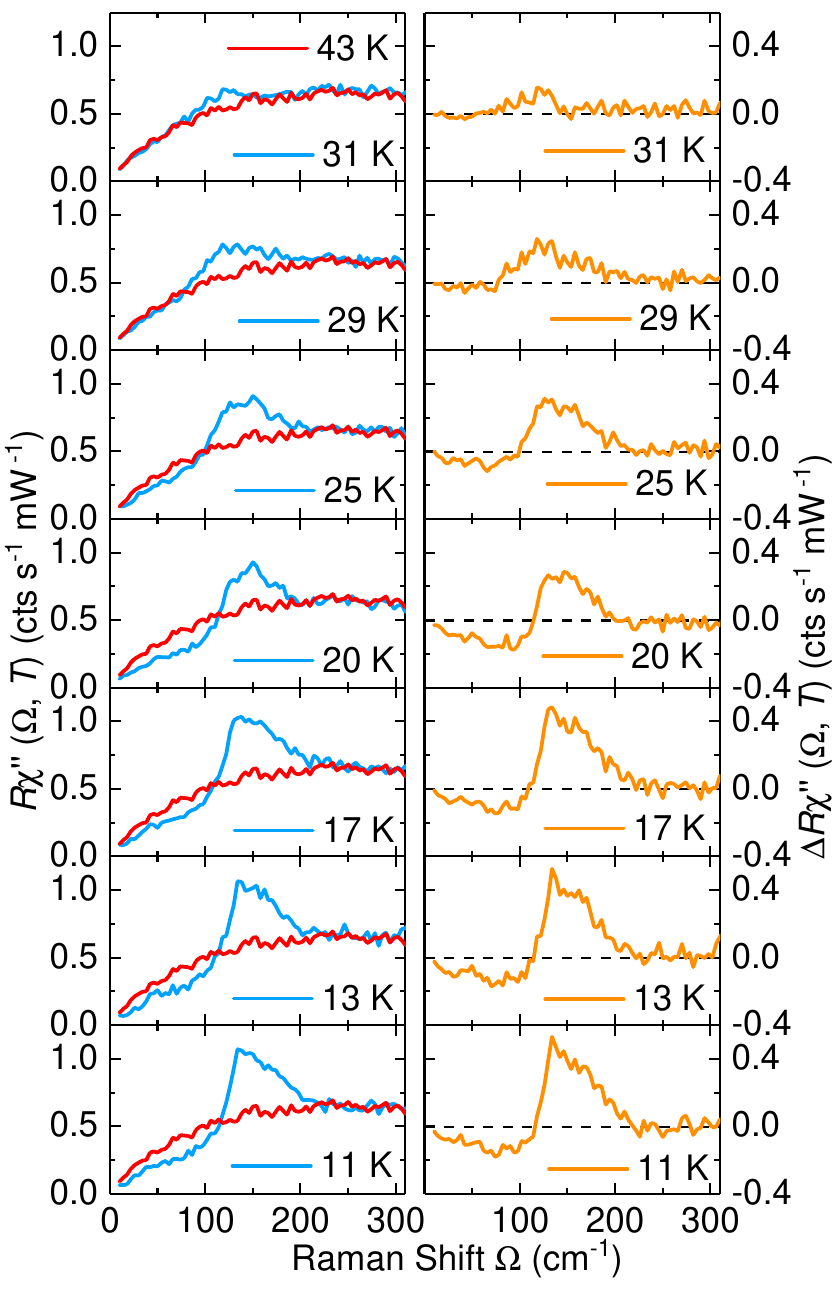}
  \caption{\Blg response of \CKFA for 43\,K and temperatures below \Tc as indicated. The raw response is shown on the left, the difference spectra are on the right. Close to 50\,cm$^{-1}$ there is a weak maximum in the superconducting state observable in the raw data and the difference spectra in the temperature range 11-20\,K.
  }
  \label{Afig:B1g}
\end{figure}

\begin{figure}[ht]
  \centering
  \includegraphics[width=7.5cm]{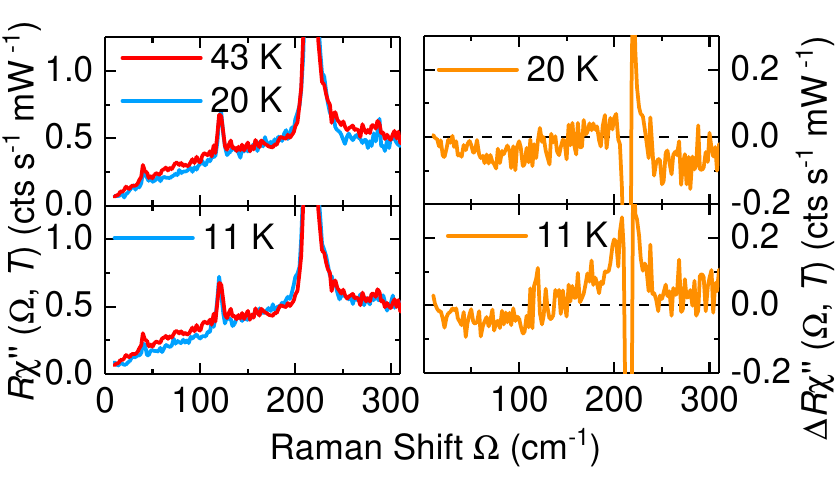}
  \caption{\BZg response of \CKFA for 43\,K and temperatures below \Tc as indicated. The raw response is shown on the left, the difference spectra are on the right.
  }
  \label{Afig:B2g}
\end{figure}

%------------------------------------------------------------------------------------------------------------
The left panels of Figs.~\ref{Afig:A1g}, \ref{Afig:B1g}, and \ref{Afig:B2g} show the raw data (after division by the Bose factor) above \Tc and at various temperatures below \Tc for \CKFA. The difference spectra according to Eq.~\eqref{eqn:differencespectrum} are shown in the right column.

%\begin{minipage}[h]{\columnwidth}
%\end{minipage}
%\vspace*{-150mm}

%\hspace*{85mm}
%\begin{minipage}[h]{\columnwidth}

%\end{minipage}

\end{appendix}

\clearpage
\bibliography{Jost_CKFA_180517}
%\bibliography{D:/!papers/!bib/literatureR2}
\end{document}